\documentclass[conference]{IEEEtran}
\IEEEoverridecommandlockouts
\usepackage{cite}
\usepackage{amsmath,amssymb,amsfonts}
\usepackage{algorithmic}
\usepackage{graphicx}
\usepackage{multirow}
\usepackage{textcomp}
\usepackage{subcaption}
\usepackage{xcolor}
\usepackage{acro}
\usepackage{tabularx}
\usepackage{makecell}
\setcellgapes{5pt}
\def\BibTeX{{\rm B\kern-.05em{\sc i\kern-.025em b}\kern-.08em
    T\kern-.1667em\lower.7ex\hbox{E}\kern-.125emX}}

\newcolumntype{L}{>{\raggedright\arraybackslash}X}
\newcolumntype{Y}{>{\centering\arraybackslash}X}

\DeclareAcronym{ML}{
  short=ML,
  long=Machine Learning
}
\DeclareAcronym{NSD}{
  short=NSD,
  long=Network Service Detector
}
\DeclareAcronym{CG}{
  short=CG,
  long=Cloud-gaming
}
\DeclareAcronym{RT}{
  short=RT,
  long=Real-time
}
\DeclareAcronym{NRT}{
  short=NRT,
  long=Non-real-time
}
\DeclareAcronym{MG}{
  short=MG,
  long=Online Mobile-gaming
}
\DeclareAcronym{VC}{
  short=VC,
  long=VOIP Video-call
}
\DeclareAcronym{AC}{
  short=AC,
  long=VOIP Audio-call
}
\DeclareAcronym{FD}{
  short=FD,
  long=File-transferring
}
\DeclareAcronym{VS}{
  short=VS,
  long=Video-streaming \& others
}
\DeclareAcronym{GBDT}{
  short=GBDT,
  long=Gradient Boosting Decision Trees
}
\DeclareAcronym{L1}{
  short=L1,
  long=First Layer
}
\DeclareAcronym{L2}{
  short=L2,
  long=Second Layer
}
\DeclareAcronym{XGBoost}{
  short=XGBoost,
  long=eXtreme Gradient Boosting
}
\DeclareAcronym{IANA}{
  short=IANA,
  long=Internet Assigned Numbers Authority
}
\DeclareAcronym{NAT}{
  short=NAT,
  long=Network Address Translation
}
\DeclareAcronym{TWT}{
  short=TWT,
  long=Target Wake Time
}
\DeclareAcronym{DPI}{
  short=DPI,
  long=Deep Packet Inspection
}
\DeclareAcronym{RSSI}{
  short=RSSI,
  long=Received Signal Strength Indicator
}
\DeclareAcronym{PID}{
  short=PID,
  long=Process ID
}
\DeclareAcronym{QoS}{
  short=QoS,
  long=Quality of Service
}
\DeclareAcronym{AI}{
  short=AI,
  long=Artificial Intelligence
}
\DeclareAcronym{UL}{
  short=UL,
  long=Uplink
}
\DeclareAcronym{DL}{
  short=DL,
  long=Downlink
}
\DeclareAcronym{IAT}{
  short=IAT,
  long=Inter-Arrival Time
}

\begin{document}

\title{Towards Intelligent Network Management: \\Leveraging AI for Network Service Detection\\
}

\author{
    \IEEEauthorblockN{
        Khuong N. Nguyen\IEEEauthorrefmark{1}, 
        Abhishek Sehgal\IEEEauthorrefmark{1}, 
        Yuming Zhu\IEEEauthorrefmark{1},
        Junsu Choi\IEEEauthorrefmark{2}, \\
        Guanbo Chen\IEEEauthorrefmark{1},
        Hao Chen\IEEEauthorrefmark{1}, 
        Boon Loong Ng\IEEEauthorrefmark{1},
        Charlie Zhang\IEEEauthorrefmark{1}
    }
    \IEEEauthorblockA{\IEEEauthorrefmark{1}Standards and Mobility Innovation Laboratory - Samsung Research America}
    \IEEEauthorblockA{\IEEEauthorrefmark{2}Samsung Electronics Co., Ltd}
    \{k.nguyen1, abhishek.s4, yuming.zhu, junsu.choi, guanbo.h, hao.chen1, b.ng, jianzhong.z\}@samsung.com
}

\maketitle

\begin{abstract}
As the complexity and scale of modern computer networks continue to increase, there has emerged an urgent need for precise traffic analysis, which plays a pivotal role in cutting-edge wireless connectivity technologies. This study focuses on leveraging Machine Learning methodologies to create an advanced network traffic classification system. We introduce a novel data-driven approach that excels in identifying various network service types in real-time, by analyzing patterns within the network traffic. Our method organizes similar kinds of network traffic into distinct categories, referred to as network services, based on latency requirement. Furthermore, it decomposes the network traffic stream into multiple, smaller traffic flows, with each flow uniquely carrying a specific service. Our ML models are trained on a dataset comprised of labeled examples representing different network service types collected on various Wi-Fi network conditions. Upon evaluation, our system demonstrates a remarkable accuracy in distinguishing the network services. These results emphasize the substantial promise of integrating Artificial Intelligence in wireless technologies. Such an approach encourages more efficient energy consumption, enhances Quality of Service assurance, and optimizes the allocation of network resources, thus laying a solid groundwork for the development of advanced intelligent networks.
\end{abstract}


\section{Introduction}
With the rapid expansion of wireless networks and a surge in the number of interconnected devices, there is a pressing necessity for advanced network management solutions to uphold optimal performance levels. Traditional network service detection methods based on port numbers and deep packet inspection, are limited in terms of scalability and effectiveness. To address these limitations, this research paper focuses on the design of a system that can detect the types of network traffic, termed as network services, using \ac{ML} techniques, with an emphasis on analyzing network traffic to detect the patterns.

Our proposed ML-based \ac{NSD} system aims to elevate both accuracy and efficiency in pinpointing services encapsulated in network traffic streams. This elevation not only aids in superior resource distribution and network regulation but also offers a detailed analysis of network traffic by scrutinizing the data exchanged between different endpoints, thereby yielding insights into the individual demands of each service.

To our knowledge, this solution represents a pioneering effort with significant potential to drive progress across a broad range of fields in both Wi-Fi and cellular technologies. The proposed system has been incorporated into our commercial products, from our flagship Galaxy S series to the Galaxy Fold and Flip devices. It addresses key concerns like traffic prioritization, \ac{QoS} assurance, and enhanced energy efficiency. By investigating the ramifications of our efforts in these sectors, we aim to underscore the transformative impact our system could have on the future development of intelligent Wi-Fi technologies.

The remainder of this paper is organized as follows to ensure a coherent explanation of our work. We initiate with a section providing a background on the relevant existing literature. This is followed by an in-depth presentation of our methodology and the system architecture designed to facilitate network service detection. Subsequently, we detail the experiments undertaken and their respective outcomes. The paper ends with a conclusion, highlighting the insights garnered through this research and proposing potential directions for future explorations.

\section{Related Work}
\begin{figure*}[ht!]
    \centering
    \includegraphics[width=0.99\textwidth]{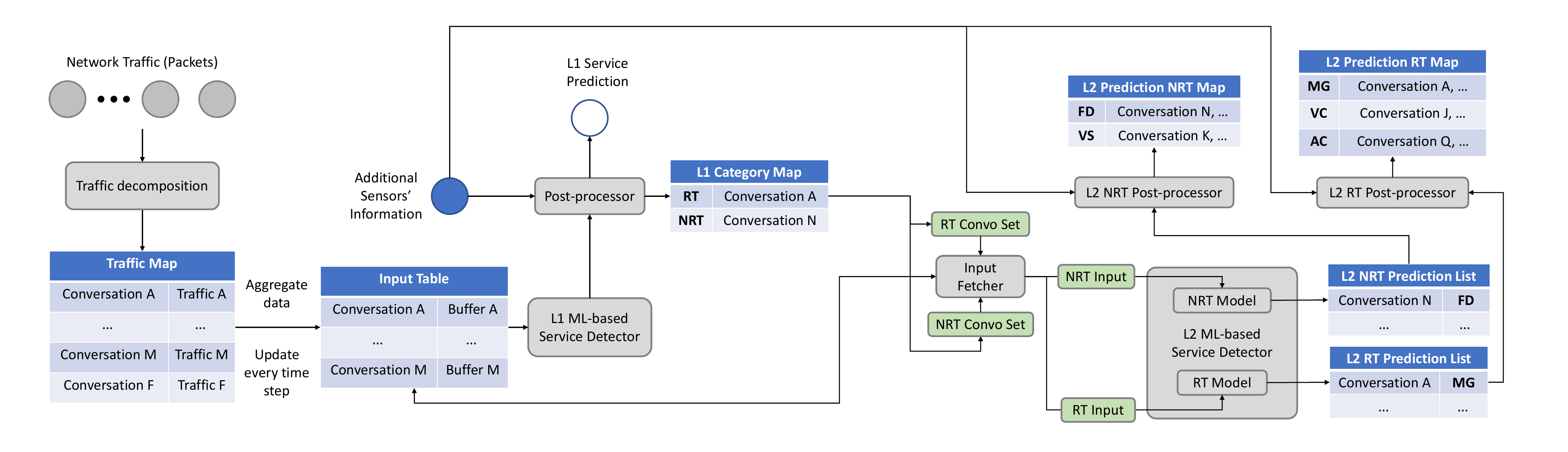} % arxiv
    \caption{The NSD system uses a multi-tiered architecture for analyzing and categorizing network traffic into distinct service types. It comprises key components such as the Traffic Decomposition Module for initial packet processing and feature extraction, the Input Management Module for assembling final input vectors, and hierarchical Machine Learning-Based Service Detectors (L1 and L2) for service categorization. These components are further refined by the Post-processors that integrate ML predictions with additional sensor data. Overall, the system aims for high accuracy and minimal error in service detection through its comprehensive and integrated approach.}
    \label{fig:sa}
\end{figure*}

Network traffic classification has garnered considerable attention in recent years, owing to its pivotal role in enhancing network management. A variety of methods have been explored in the literature, each with its unique strengths and weaknesses. Three primary approaches dominate the field: port-based methods, packet inspection techniques, and statistical analysis.

The port-based approach is among the oldest and most straightforward methods for traffic classification. It relies on the extraction of port numbers from TCP/UDP headers, which are then matched against the \ac{IANA} standard port numbers for traffic identification \cite{moore2005toward, dainotti2012issues}. While simple and quick, this method suffers from several limitations. Due to port obfuscation, \ac{NAT}, and other techniques that alter/hide port numbers, the effectiveness of this approach has declined.

Packet inspection techniques, also known as \ac{DPI}, primarily target the application-layer payload within data packets, employing predefined patterns to recognize network protocols \cite{yeganeh2012cute}. Despite their utility, \ac{DPI} approaches encounter challenges, including concerns over user privacy and the necessity for regular updates to their pattern libraries. Recently, Lotfollahi et al. presented a novel framework called Deep Packet. This framework leverages deep learning autoencoder algorithms to automatically extract features from captured network packets, thereby facilitating the classification of encrypted traffic \cite{lotfollahi2020deep}.

Statistical analysis methods provide a nuanced avenue for understanding and categorizing network traffic. Initial methodologies leveraging statistical techniques focused on attributes like packet inter-arrival times and packet sizes for traffic identification \cite{crotti2007traffic}. Over time, these methods have increasingly incorporated \ac{AI} by utilizing \ac{ML} algorithms to manage large datasets effectively. These ML-based strategies have been applied in various contexts, utilizing a gamut of algorithms from unsupervised methods like k-nearest neighbors to supervised deep learning neural networks \cite{yamansavascilar2017application, wang2015applications}. Qiu et al. present a solution for network service detection that employs the generated predictions to configure the \ac{TWT}. However, it is important to note that the operational scope of their system is limited to traffic streams carrying a single service \cite{qiu2021category}. Overall, such techniques often necessitate expert-guided feature selection and extraction, rendering the process both time-consuming and susceptible to human error.

 In other work \cite{holland2021new, holland2022towards}, network traffic is standardized and analyzed through feature extraction from packet headers and payloads. While effective for detecting intrusions and network anomalies, this approach overlooks latency and throughput requirements. Moreover, its reliance on individual packet features risks misclassification due to future protocol changes.

\section{Methodology}
In this section, we outline the approach that forms the basis for developing and implementing the proposed ML-based \ac{NSD} system.

\subsection{Defining The Network Service and Categories}
Classifying network traffic is synonymous with identifying the type of service carried by that traffic. A Network Service is a function provided over a network infrastructure that facilitates application-level interactions and data exchanges between connected devices. Network services are organized into three primary distinct categories based on specific latency requirements: very low latency, low latency, and high latency. In this paper, the terms network traffic classification and network service detection are occasionally used interchangeably.

The categorization of these categories is informed by the distinct needs of applications that fall into these latency brackets. Notably, applications demanding exceptionally low latency are typically associated with cloud gaming experiences. This led us to label the category necessitating very low latency as \ac{CG} (lower than 50 ms) for application such as XBox Game Pass, Google Stadia, etc.

To define the categories of low and high latency, we capitalized on the interaction characteristics inherent to each application. Specifically, applications that thrive on frequent user interaction or substantial \ac{UL}/\ac{DL} exchanges, such as online mobile gaming (e.g., Call of Duty, Fortnite, etc.) and VOIP calls (e.g., WhatsApp, Google Meet, etc.), inherently demand relatively low latency. Conversely, applications with less interaction, like video/audio streaming (e.g. Netflix, YouTube, Spotify, etc.), File-transferring (e.g., link downloading, Dropbox, etc.) or web browsing, exhibit more lenient latency requirements. This inherent attribute of interactivity, where interactive applications maintain bi-directional traffic while non-interactive ones display a clear dominance in one traffic direction, aids in distinguishing these categories. We have named the category with substantial interaction as \ac{RT} service type (latency between 50ms and 200ms is tolerable), while the counterpart with less real-time demands is termed \ac{NRT} service type (latency less than 500ms is preferable).

To enhance the granularity of our classification, we delve deeper into the \ac{RT} and \ac{NRT} service categories, breaking them down into more refined sub-categories. Within the \ac{RT} service category, we describe three distinct sub-categories, encompassing \ac{MG}, \ac{VC}, and \ac{AC} services. Similarly, within the \ac{NRT} service category, we establish two distinct sub-categories: \ac{FD} service and \ac{VS} service. This refined classification framework enables a more nuanced understanding of the diverse range of network services and their associated characteristics.

\subsection{Network Traffic Tracking and Feature Selection}\label{med:ttfs}
To classify network traffic, it is essential to capture and analyze network packets in real-time. For this purpose, we design a polling mechanism that operates at 500-millisecond (1 time-step) intervals to gather all packets transmitted within that time frame. Subsequently, we parse the IP headers of these packets to extract pertinent data. Utilizing this data, we compute a set of 10 statistical features within a single time step, as follows:
\begin{itemize}
    \item \ac{UL} Maximum \ac{IAT}: This represents the maximal temporal difference between the arrivals of consecutive packets (1 value).
    \item \ac{UL} Average \ac{IAT}: This denotes the average time difference between the arrivals of consecutive packets (1 value).
    \item \ac{UL} and \ac{DL} Packet Counts: These are the total counts of packets sent in both the \ac{UL} and \ac{DL} directions (2 values).
    \item \ac{UL} and \ac{DL} Minimum Packet Size: These signify the smallest packet sizes, measured in megabytes, for both \ac{UL} and \ac{DL} traffic (2 values).
    \item \ac{UL} and \ac{DL} Maximum Packet Size: These indicate the largest packet sizes, again measured in megabytes, for both \ac{UL} and \ac{DL} traffic (2 values).
    \item \ac{UL} and \ac{DL} Average Packet Size: These features reflect the mean packet sizes, in megabytes, for \ac{UL} and \ac{DL} traffic (2 values).
\end{itemize}
These features are used for the service classification purpose.

\subsection{Network Traffic Decomposition}
\begin{figure}[ht!]
    \centering
    \includegraphics[width=0.45\textwidth]{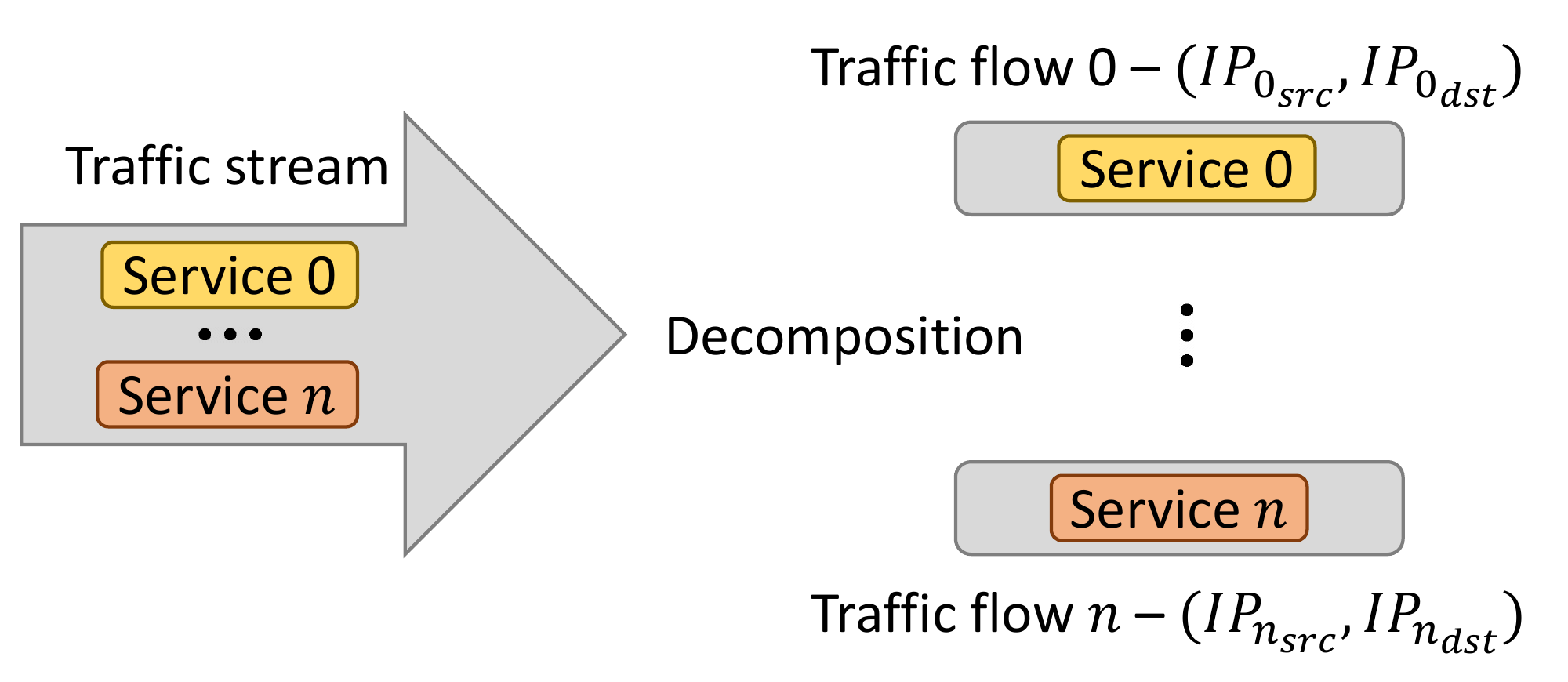} % arxiv
    \caption{To accurately identify all services within a network stream, it is essential to decompose it into multiple smaller traffic flows, each carrying a distinct network service.}
    \label{fig:tdc}
\end{figure}

A single network traffic stream may encompass multiple services. To accurately identify all services within such a stream, it is imperative to decompose it into multiple smaller traffic flows, each carrying a distinct service.

To decompose a network traffic stream into individual flows, we employ the quintuple rule for packet grouping. The quintuple rule utilizes a five-tuple set—comprising the Source IP Address, Source Port Number, Destination IP Address, Destination Port Number, and Protocol (e.g., TCP, UDP)—to uniquely identify a specific session or flow within the network traffic. By leveraging these five parameters collectively, we can accurately identify and manage discrete flows or sessions of data packets. This technique is useful in various network contexts, including routing, firewall policy enforcement, and traffic monitoring, and serves as a foundational element in network service detection.

Specifically, we define a traffic flow by utilizing a tuple consisting of the Source and Destination IP addresses, extracted from the packets' IP headers. Packets are then grouped based on this tuple, which we refer to as a conversation to denote the ongoing communication between two endpoints.

\subsection{The Machine Learning Method}
\begin{figure}[ht!]
    \centering
    \includegraphics[width=0.4\textwidth]{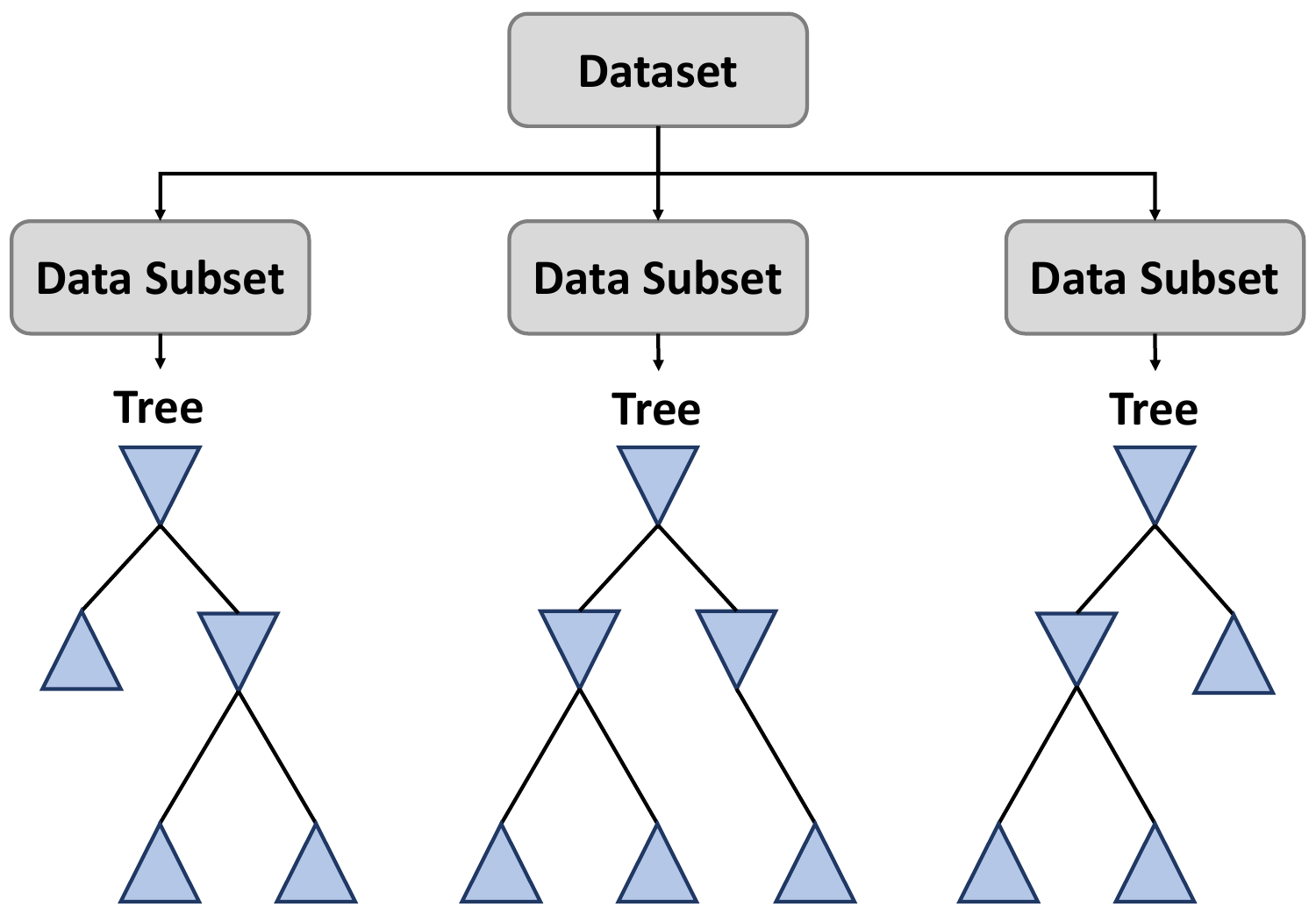} %  arxiv
    \caption{XGBoost belongs to the gradient boosting family of algorithms, which iteratively improves the predictive accuracy of a model by combining the strengths of multiple weak learners, typically decision trees.}
    \label{fig:xgb}
\end{figure}

XGBoost, or eXtreme Gradient Boosting \cite{chen2016xgboost}, is a highly acclaimed \ac{ML} algorithm renowned for its exceptional predictive capabilities and versatility. As a member of the gradient boosting family, XGBoost iteratively enhances model predictions by leveraging the collective strength of multiple weak learners, typically in the form of decision trees. Its distinct advantage lies in its aptitude for handling structured, heterogeneous, and tabular data, rendering it a favored choice across a spectrum of practical applications.

Our adoption of XGBoost as the primary methodology in this research is grounded in its empirical superiority over other \ac{ML} techniques, including neural networks, when confronted with tasks that involve structured, heterogeneous, or tabular data. \ac{GBDT} algorithms \cite{natekin2013gradient}, like XGBoost, have consistently exhibited their efficacy in diverse domains such as finance, healthcare, wireless communication, and recommendation systems. This robust performance stems from their ability to capture intricate data relationships, effectively manage missing data, and furnish feature importance metrics, thus enhancing model interpretability. Moreover, XGBoost offers a gamut of optimization techniques, including regularization and parallel processing, fortifying its performance and scalability. Hence, the rationale behind our selection of XGBoost lies in its proven track record of surmounting the challenges posed by structured data analysis, making it an optimal choice for addressing this problem.

\subsection{Data Collection}
\begin{figure}[h!]
    \centering
    \includegraphics[width=0.45\textwidth]{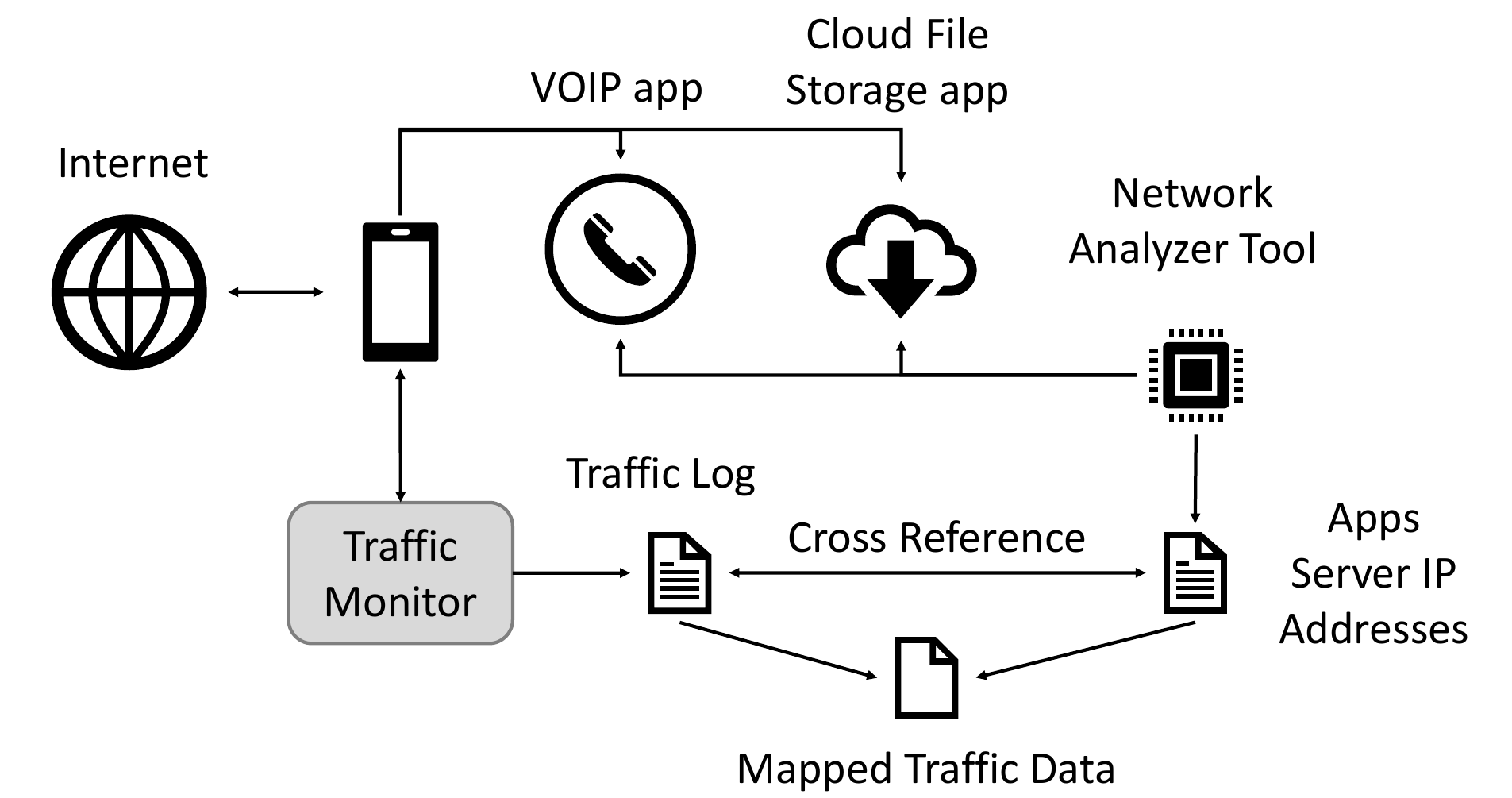} % arxiv
    \caption{The figure illustrates our data extraction and labeling method, which involves associating network traffic data with specific applications and network services. This process includes capturing active applications (retrieving Process IDs), employing network analyzers to monitor connections \& extract server IPs, and integrating this information with traffic logs.}
    \label{fig:dc}
\end{figure}

To effectively train and evaluate a model for predicting service types based on network traffic data, it is imperative to follow a systematic process for decomposing and mapping this data to the relevant applications/packages. The following outlines a series of essential steps to achieve this intricate task.

The initial step involves capturing a comprehensive list of currently running applications or packages that are actively generating network traffic. This can be achieved by monitoring the system's processes and identifying those responsible for network data transmission. Subsequently, it is crucial to obtain the \ac{PID}s associated with each of the identified applications/packages. \ac{PID}s serve as unique identifiers for processes running on the system and are pivotal for tracking network activity at the process level. Employing specialized network analyzer tools becomes instrumental in the next phase. Tool such as Netstat is utilized to track the network system calls and active connections attributed to each \ac{PID}. By doing so, a granular understanding of the network traffic generated by each application/package is attained. Within the network analyzer tools, the system's active connections are examined to extract the relevant server IP addresses. These IP addresses represent the destinations with which the applications/packages are communicating over the network. This information is pivotal for associating network traffic with specific external servers. The information acquired from the network analyzer tools is then integrated with the traffic log data. This log typically contains records of network activity, including timestamps, data volumes, and source IP addresses. By cross-referencing the server IP addresses collected from the network analyzing tool with the data in the traffic log, it becomes possible to precisely identify which data corresponds to specific applications/packages. A list of applications that were used to generate traffic data is provided in Table \ref{tab:al}.

\begin{table}[htbp!]
    \centering
    \begin{tabularx}{.48\textwidth}{|c|Y|Y|} \hline
        Cloud-gaming (CG)& Real-time (RT)& Non-real-time (NRT)\\ \hline \hline
        XBox Game Pass & \makecell{Facebook Messenger\\(VC \& AC)} & \makecell{Samsung Internet\\Link DL (FD)}\\ \hline
        Google Stadia & \makecell{Skype\\(VC \& AC)} & \makecell{Google Chrome\\Link DL (FD)}\\ \hline
        Netboom & \makecell{Webex\\(VC \& AC)} & \makecell{Google Drive \\DL \& UL (FD)}\\ \hline
        Mogul & \makecell{Discord\\(VC \& AC)} & \makecell{DropBox\\DL \& UL (FD)}\\ \hline
        Geforce Now & \makecell{Telegram\\(VC \& AC)} & \makecell{Amazon Drive\\DL \& UL (FD)}\\ \hline
        Amazon Luna & \makecell{Viber\\(VC \& AC)} & \makecell{Netflix\\(VS)}\\ \hline
        & \makecell{Whatsapp\\(VC \& AC)} & \makecell{Disney+\\(VS)}\\ \hline
        & \makecell{Google Duo\\(VC \& AC)} & \makecell{Hulu\\(VS)}\\ \hline
        & \makecell{Google Meet\\(VC \& AC)} & \makecell{ESPN+\\(VS)}\\ \hline
        & \makecell{Zoom\\(VC/AC)} & \makecell{YouTube\\(VS)}\\ \hline
        & \makecell{Microsoft Teams\\(VC \& AC)} & \makecell{Samsung Internet\\(VS)}\\ \hline
        & \makecell{PUBG\\(MG)} & \makecell{Google Chrome\\(VS)}\\ \hline
        & \makecell{Among Us\\(MG)} & \\ \hline
        & \makecell{Call of Duty\\(MG)} & \\ \hline
        & \makecell{League of Legends\\(MG)} & \\ \hline
        & \makecell{Brawlstars\\(MG)} & \\ \hline
        & \makecell{Free Fire\\(MG)} & \\ \hline
        & \makecell{Onmyoji Arena\\(MG)} & \\ \hline
    \end{tabularx}
    \caption{List of mobile applications that were used to generate traffic data and collected in our experiment. VC \& AC stands for VOIP Video-call and Audio-call. MG stands for Online-mobile-gaming. FD stands for File-transferring. VS stands for Video-streaming and Others.}
    \label{tab:al}
\end{table}

We gathered data under various Wi-Fi conditions to encompass a comprehensive range of scenarios. These conditions included different Received Signal Strength Indicator \ac{RSSI} levels, with normal levels being $\geq -55 \text{ dBm}$ and edge levels at $\leq -65 \text{ dBm}$. We also considered different levels of traffic contention, categorizing them as normal ($\text{cca/radio\_on}< 0.1$), mildly congested ($0.2 \leq \text{cca/radio\_on} \leq 0.4$), and highly congested ($\text{cca/radio\_on} > 0.55$). Additionally, data were collected across multiple Wi-Fi bands, specifically 2.4 GHz, 5 GHz, and 6 GHz.

\section{System Architecture}
The \ac{NSD} system proposed in this study is a sophisticated architecture designed to track and analyze network traffic packets while segregating them into multiple streams based on conversations. The system architecture (described in fig. \ref{fig:sa}) can be divided into several key components, including The traffic decomposition module, the input management module, the \ac{L1} and \ac{L2} ML-based service detectors, and their corresponding post processors. These component are described as follows:

The Traffic Decomposition Module functions as the initial processing unit is responsible for decomposing network traffic into distinct conversations while extracting relevant features. Specifically, this module monitors all packets within a designated 500 millisecond interval, referred to as a time step and computes ten statistical features, as described in Section \ref{med:ttfs}. Packet IP headers are parsed to extract source and destination IP addresses, which are then utilized to construct individual conversations. Concurrently, a dynamic traffic map is maintained to correlate these computed features with their corresponding conversations during each time step. To enhance the reliability of the data, a filtering mechanism is used to exclude irrelevant conversations, such as those associated with broadcast IP addresses, which are not contributory to the predictive analytics of network services. One advantage of employing this methodology is its capability to accurately identify mixed services within a single network traffic stream.

\begin{figure}[ht!]
    \centering
    \includegraphics[width=0.48\textwidth]{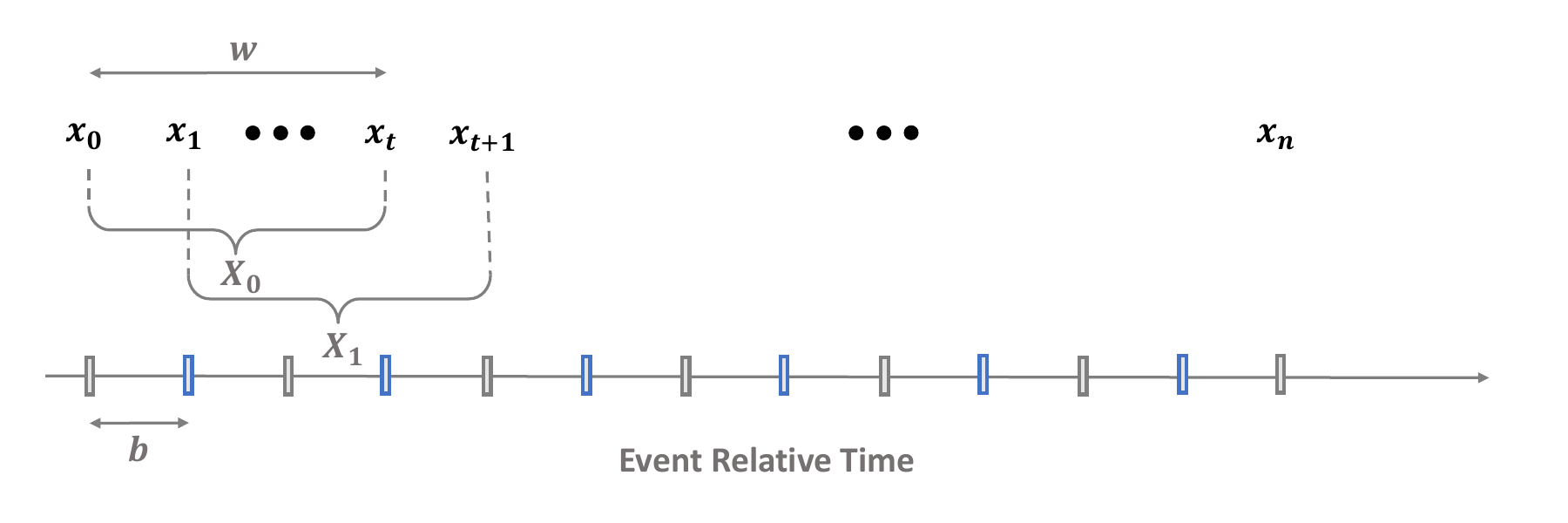} % for ieee
    \caption{the process to form the input for each conversation can be thought of as to slide a window of size $w=6$ over the features’ series of that conversation. Each time step is $b=500$ ms. At time $t$, the input $X_t$ consist of a combination of 6 feature vectors $[x_{t-w+1},…,x_t]$. Therefore, the input at time $t$, $X_t$ comprises of the following feature vectors $[x_{t-5},x_{t-4},x_{t-3},x_{t-2},x_{t-1},x_t]$.}
    \label{fig:if}
\end{figure}

The Input Management Module acts as an intermediary, responsible for formulating the final inputs for the ML-based service detector. This module ingests the traffic map generated by the Traffic Decomposition Module and oversees an input table with a size of $N=7$, structured to aggregate feature data for the downstream task. Employing a moving-window approach over a span of $M=3$ seconds and encapsulating $M*2$ individual time steps, the module generates an input vector comprising of $M*2*10$ distinct variables ($3*2*10=60$ are used in our system) (Fig. \ref{fig:if}). The decision to adopt a 3-second window is supported by an empirical study, which has shown that this duration provides sufficient information for generating reliable predictions while minimizing latency at the initiation phase. At every time step, each entry from the aforementioned traffic map is mapped to a corresponding input buffer, each with a size limit of $M*2$ slots. These buffers function as dynamic queues, storing input features in a time-ordered sequence where the most recent conversation resides at the head and the oldest at the tail. To manage the input table capacity, the oldest conversations are automatically evicted to make room for new entries. Additionally, to compensate for conversations with no new incoming traffic data, the module injects dummy traffic chunks, thereby ensuring a complete and uninterrupted data stream.

The ML-Based Service Detectors serve as the computational nucleus of the system, tasked with processing input vectors from the Input Management Module to generate service-type predictions for corresponding conversations at 500-millisecond intervals. The architecture is hierarchical, featuring two layers for service detection: the \ac{L1} employs a single trained XGBoost model to classify services into three primary categories \ac{CG}, \ac{RT}, and \ac{NRT}. The \ac{L2} contains two specialized XGBoost models; one identifies \ac{RT} service sub-categories, namely, \ac{MG}, \ac{VC}, and \ac{AC}, while the other discerns \ac{NRT} service sub-categories like \ac{FD} and \ac{VS}. Prior to model inference, a buffer size validation step is executed for each entry within the input table; entries that do not meet the minimum buffer size requirement of $M$ time steps (default is 6) are systematically ignored to ensure the reliability and integrity of the predictive process. The initial coarse-grained service predictions are generated by \ac{L1} and organized into a data structure termed the \ac{L1} Category Map. These are subsequently consolidated into a structured multi-label output featuring three distinct fields corresponding to \ac{CG}, \ac{RT}, and \ac{NRT}, presented in this specific order. Each field within the multi-label output is conditionally activated when its corresponding service type is accurately identified by the prediction algorithm. This structured output a standardized format serves to provide a comprehensive and orderly representation of the identified services within a traffic stream. Finally, \ac{L2} utilizes the \ac{L1} predictions for further service sub-categorization; \ac{RT}-classified conversations are processed by the \ac{L2} \ac{RT} model, and \ac{NRT}-classified conversations by the \ac{L2} \ac{NRT} model. The multi-layered architecture aims to minimize service detection errors, constraining the impact of inaccuracies even if they occur at the second layer.

The last elements in the pipeline are the Post-processors, designed to integrate the preliminary predictions derived from the preceding ML-based service detectors with supplementary sensor data to generate refined final outputs. There are a total of three post-processors. One for the \ac{L1} and two for the \ac{L2}. Each of these post-processors employ a queue-like data structure, known as the historical prediction buffer, which has a capacity of 7 slots for storing predictions rendered at each time step. During each computational cycle, a composite output is formulated through a majority-voting mechanism applied to the historical prediction buffer. In addition to this, sensor-derived attributes are strategically incorporated to refine and expedite the final service-type prediction. For instance, one such attribute is the gaming flag provided by Samsung's Game Optimizing Service in the Android system, which signifies whether the foreground application is a mobile game, a strong indicator that the current service type is \ac{RT}. Another sensor input of note pertains to camera usage: activation of either the front or back camera, combined with a sufficient number of \ac{RT} service predictions in the historical prediction buffer, serves as a robust indicator that the user is engaged in a VOIP video call, which falls under the \ac{RT} service category.

In summary, the \ac{NSD} system epitomizes a multi-tiered, integrated approach to analyze the network service. Through an assembly of sophisticated modules, each performing specialized tasks ranging from traffic decomposition to predictive analytics, the system offers a comprehensive solution for accurately identifying and classifying multiple types of network services.

\section{Experimental Results}

The evaluation of the \ac{NSD} system's performance was conducted using key metrics such as accuracy, precision, recall, and F1-score. A diverse test dataset, encompassing various types of network traffic and different Wi-Fi bands, was employed. The system exhibited high performance, achieving an overall accuracy rate of 98.8\% for \ac{L1}, 97.5\% for \ac{L2} \ac{RT}, and 90.1\% for \ac{L2} \ac{NRT}. These high scores affirm the system's ability to accurately categorize diverse network services.

\begin{table}[htbp!]
    \centering
    \begin{tabular}{|c|c|c|c|c|} \hline
         & Precision & Recall & F1-score & Support\\ \hline \hline
        \multicolumn{5}{|c|}{2.4 GHz}\\ \hline
        Cloud-gaming (CG) & 0.98 & 1.00 & 0.99 & 40,900\\ \hline
        Real-time (RT) & 1.00 & 0.98 & 0.99 & 152,609\\ \hline
        Non-real-time (NRT) & 0.97 & 0.99 & 0.98 & 70,532\\ \hline
        Accuracy &  &  & 0.99 & 264,041\\ \hline \hline
        \multicolumn{5}{|c|}{5 GHz}\\ \hline
        Cloud-gaming (CG) & 1.00 & 1.00 & 1.00 & 111,972\\ \hline
        Real-time (RT) & 0.99 & 0.99 & 0.99 & 448,815\\ \hline
        Non-real-time (NRT) & 0.98 & 0.98 & 0.98 & 182,690\\ \hline
        Accuracy &  &  & 0.99 & 743,477\\ \hline \hline
        \multicolumn{5}{|c|}{6 GHz}\\ \hline
        Cloud-gaming (CG) & 1.00 & 1.00 & 1.00 & 58,987\\ \hline
        Real-time (RT) & 0.99 & 0.99 & 0.99 & 235,711\\ \hline
        Non-real-time (NRT) & 0.98 & 0.98 & 0.97 & 86,475\\ \hline
        Accuracy &  &  & 0.99 & 381,173\\ \hline
    \end{tabular}
    \caption{L1 classification report for different Wi-Fi bands.}
    \label{tab:l1cr}
\end{table}

\begin{figure}[ht!]
    \centering
    \includegraphics[width=0.4\textwidth]{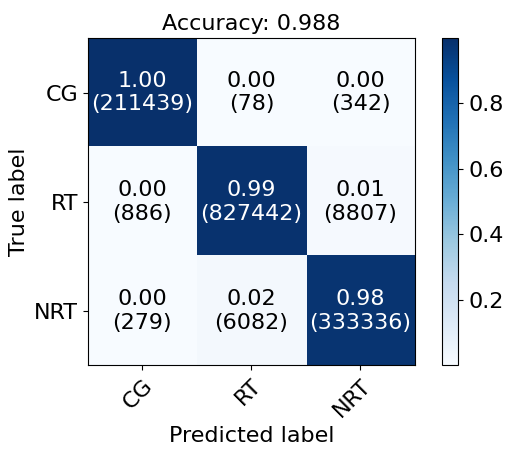} % arxiv
    \caption{L1 confusion matrix.}
    \label{fig:l1cm}
\end{figure}

Subsequent discussions present in-depth results for \ac{L1}. The confusion matrix for \ac{L1}, depicted in Fig. \ref{fig:l1cm}, further validates the system's ability. It shows remarkably low rates of false positives and false negatives across classes, thereby confirming its effectiveness in coarse-grained service classification tasks. A detailed classification report is presented in Table \ref{tab:l1cr}, where high values for precision, recall, and F1-score were observed across all service categories (CG, RT, and NRT). These multifaceted metrics substantiate the system's robustness and appropriateness for real-world application deployment.

\begin{table}[htbp!]
    \centering
    \begin{tabular}{|c|c|c|c|c|} \hline
         & Precision & Recall & F1-score & Support\\ \hline \hline
        \multicolumn{5}{|c|}{2.4 GHz}\\ \hline
        Mobile-gaming (MG) & 0.95 & 0.99 & 0.97 & 25,832\\ \hline
        Video-call (VC) & 0.99 & 0.95 & 0.97 & 37,414\\ \hline
        Audio-call (AC) & 0.95 & 0.96 & 0.96 & 36,800\\ \hline
        Accuracy &  &  & 0.97 & 100,046\\ \hline \hline
        \multicolumn{5}{|c|}{5 GHz}\\ \hline
        Mobile-gaming (MG) & 0.96 & 1.00 & 0.98 & 82,020\\ \hline
        Video-call (VC) & 0.99 & 0.97 & 0.98 & 154,642\\ \hline
        Audio-call (AC) & 0.97 & 0.97 & 0.97 & 143,023\\ \hline
        Accuracy &  &  & 0.98 & 379,685\\ \hline \hline
        \multicolumn{5}{|c|}{6 GHz}\\ \hline
        Mobile-gaming (MG) & 0.98 & 1.00 & 0.99 & 39,474\\ \hline
        Video-call (VC) & 1.00 & 0.95 & 0.98 & 90,364\\ \hline
        Audio-call (AC) & 0.96 & 1.00 & 0.98 & 90,046\\ \hline
        Accuracy &  &  & 0.98 & 219,884\\ \hline
    \end{tabular}
    \caption{L2 Real-time classification reports for different Wi-Fi bands.}
    \label{tab:l2rtcr}
\end{table}

\begin{table}[htbp!]
    \centering
    \begin{tabular}{|c|c|c|c|c|} \hline
         & Precision & Recall & F1-score & Support\\ \hline \hline
        \multicolumn{5}{|c|}{2.4 GHz}\\ \hline
        File-transferring (FD) & 0.74 & 0.98 & 0.85 & 2,325\\ \hline
        Video-streaming (VS) & 0.98 & 0.68 & 0.80 & 2,464\\ \hline
        Accuracy &  &  & 0.83 & 4,789\\ \hline \hline
        \multicolumn{5}{|c|}{5 GHz}\\ \hline
        File-transferring (FD) & 0.86 & 0.94 & 0.90 & 5,282\\ \hline
        Video-streaming (VS) & 0.91 & 0.79 & 0.85 & 3,898\\ \hline
        Accuracy &  &  & 0.88 & 9,180\\ \hline \hline
        \multicolumn{5}{|c|}{6 GHz}\\ \hline
        File-transferring (FD) & 1.00 & 0.98 & 0.99 & 5,672\\ \hline
        Video-streaming (VS) & 0.92 & 0.98 & 0.95 & 1,076\\ \hline
        Accuracy &  &  & 0.98 & 6,748\\ \hline
    \end{tabular}
    \caption{L2 Non-real-time classification reports for different Wi-Fi bands.}
    \label{tab:l2nrtcr}
\end{table}

\begin{figure}[t]
	\centering
	% \begin{subfigure}{0.23\textwidth} % icc
        \begin{subfigure}{0.4\textwidth} % arxiv
		\centering
		\includegraphics[width=1\textwidth]{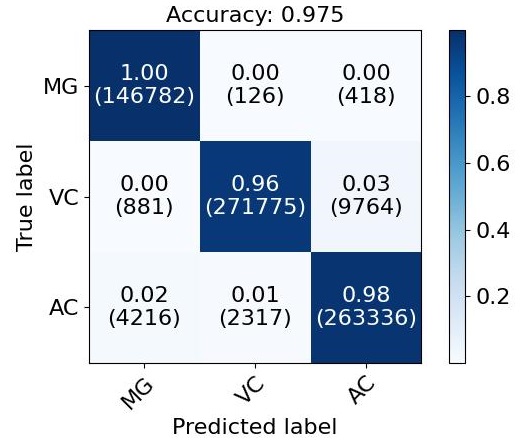}
		\caption{L2 RT.}
		\label{fig:l2rtcm}
	\end{subfigure}
	% \begin{subfigure}{0.23\textwidth} % icc
        \begin{subfigure}{0.4\textwidth} % arxiv
		\centering
		\includegraphics[width=1\textwidth]{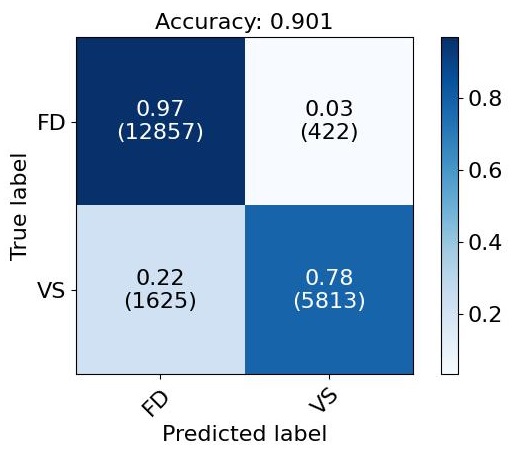}
		\caption{L2 NRT.}
		\label{fig:l2nrtcm}
	\end{subfigure}
	\caption{L2 confusion matrices.}
	\label{fig:l2cm}
\end{figure}

The fine-grained performance results for \ac{L2} are specified in Table \ref{tab:l2rtcr} for \ac{RT} services and Table \ref{tab:l2nrtcr} for \ac{NRT} services. The \ac{L2} \ac{RT} model demonstrates accuracy levels exceeding 95\%, underlining its capability to discern various \ac{RT} service sub-categories. This is further corroborated by the confusion matrix in Fig. \ref{fig:l2rtcm}. Conversely, the \ac{L2} \ac{NRT} model exhibits lower performance, yet maintains an accuracy rate above 90\%. According to the classification report in Table \ref{tab:l2nrtcr} and the confusion matrix in Fig. \ref{fig:l2nrtcm}, the \ac{L2} \ac{NRT} model excels in identifying the \ac{FD} service sub-category but shows lower accuracy for the \ac{VS} service sub-category.

In summary, the experimental results validate the \ac{NSD} system as a highly efficient, practical, and reliable mechanism for service detection. Its capacity for real-time, accurate service identification positions it as an invaluable asset for targeted applications in the realm of intelligent Wi-Fi management.

\section{Conclusion}
In conclusion, this study has delved into the realm of network traffic analysis and classification, responding to the urgent needs generated by the increasingly complex and expansive nature of modern computer networks. Through the use of \ac{ML} methodologies, we have introduced an innovative data-driven approach that excels in the precise identification of various network service types based on requirements such as latency. Our method, grounded in the analysis of network traffic patterns, effectively categorizes similar network traffic into distinct classes, termed network services.

The strength of our approach is rooted in its dual capabilities: first, the decomposition of a network traffic stream into multiple flows, each corresponding to a distinct service, and second, the real-time detection of these services, offering substantial utility for real-world applications. These capabilities are realized through a meticulously engineered system that leverages \ac{ML} models. These models have been trained on a comprehensive dataset, which includes labeled samples representing a wide array of network service types under varying Wi-Fi network conditions. Upon testing and evaluation, our system has exhibited high performance in identifying network services. Such results not only attest to the system's robustness but also highlight the immense potential for the integration of \ac{AI} within wireless technologies, especially in the specialized field of network service detection and classification.

By leveraging the power of \ac{AI}, we pave the way for more efficient energy utilization, enhanced \ac{QoS} assurance, and the optimization of network resource allocation. These outcomes collectively lay a solid foundation for the development of genuinely intelligent networks, poised to meet the evolving demands of our interconnected world. As we continue to navigate the intricacies of modern network ecosystems, the insights and methodologies presented in this study offer invaluable contributions towards the realization of smarter and more responsive wireless connectivity technologies.

\bibliographystyle{IEEEtran}
\bibliography{bibliography}{}

\end{document}